\documentstyle[12pt]{article}
\begin{document}
\title{QUANTUM MECHANICAL BLACK HOLES:TOWARDS A UNIFICATION OF
QUANTUM MECHANICS AND GENERAL RELATIVITY}
\author{B.G. Sidharth\\Centre for Applicable Mathematics \&
Computer Sciences\\
B.M. Birla Science Centre\\ Adarsh Nagar, Hyderabad - 500 063(India)\\
Dedicated to the memory of my parents}
\date{}
\maketitle
\begin{abstract}
In this paper, starting from vortices we are finally lead to a treatment
of Fermions as Kerr-Newman type Black Holes wherein we identify the
horizon at the particle's Compton wavelength periphery. A naked
singularity is avoided and the singular processes inside the horizon
of the Black Hole are identified with Quantum Mechanical effects within
the Compton wavelength.\\
Inertial mass, gravitation, electromagnetism and even QCD type interactions
emerge from such a description including relative strengths and also
other features like the anomalous gyromagnetic ratio, the discreteness
of the charge, the reason why the electron's field emerges from Newman's
complex transformation in General Relativity, a rationale for the
left handedness of neutrinos and the matter-antimatter imbalance.\\
This model describes the most fundamental stable Fermions viz., the
electrons, neutrinos and approximately the quarks. It also harmoniously
unifies the hydrodynamical, monopole and classical relativistic
perspectives.
\end{abstract}
\section{Introduction}
Ordinary Quantum Mechanics works at distances much greater than the
Compton wavelength of elementary particles or roughly $10^{-12}cm$. In the
domain of Quantum Field Theory, particles are points, space-time is a
continuum and special relativity holds, though very recently there has
been a school of thought (the spirit of Effective Field Theories) that Field Theory itself is a low energy
approximation. On the other hand in Quantum Gravity we attempt to deal
with phenomena at distances of the order of the Planck length or
$10^{-33}cm$, though there has as of now been no successful unification
of Quantum Mechanics and General Relativity.\\
In this preliminary communication, we consider an alternative viewpoint,
dealing with distances of the order of the Compton wavelengh. At this
level Quantum Mechanical phenomena like zitterbewegung and negative
energies and luminal velocities come in. Taking a route through
relativistic vortices, monopoles and classical considerations, we are lead
to the model of leptons (and also approximately quarks) as "Quantum
Mechanical Black Holes" (QMBH in what follows), wherein features of Quantum
Mechanics and General Relativity are inextricably inter-woven. At the same time,
we can trace the origin of inertial mass, gravitation, electromagnetism
and even QCD type interactions in such a picture.\\
In section 2 we invoke the DeBroglie-Bohm Hydrodynamical Formulation to
picture an elementary particle as a relativistic vortex from which it is
possible to recover its mass and quantized spin. Taking the cue from here
in section 3 we argue that the inertial mass of an elementary particle
is the energy of binding of nonlocal amplitudes in the zitterbewegung
Compton wavelength region. In section 4 it is shown how the Dirac monopole
theory is really identical to the picture of a particle as a relativistic
vortex. In section 5 the zitterbewegung is examined in greater detail and
it is argued that the usual positive energy states we encounter in the
physical universe are at scales greater than the Compton wavelength. In
section 6 it is shown how the preceding Quantum Mechanical considerations
can be equally well described in classical terms, that is for a
relativistic collection, or for a hydrodynamical flow. In section 7 it is
suggested that an electron, for example could be described by the Kerr-Newman
metric, while a full General Relativistic rationale for such an
identification is given in section 8. Finally in section 9 a number of
comments are made.\\
Ultimately there is a convergence of the various
approaches and a harmonious unified picture appears to emerge.
\section{The Bohm Hydrodynamical Formulation}
In the Bohm hydrodynamical formulation\cite{r1,r2},
we start with the Schrodinger equation
\begin{equation}
\imath \hbar \frac{\partial \psi}{\partial t} = - \frac{\hbar^2}{2m}
\nabla^2 \psi + V \psi\label{e1}
\end{equation}
In (\ref{e1}), the substitution
\begin{equation}
\psi = Re^{\imath S}\label{e2}
\end{equation}
where $R$ and $S$ are real functions of $\vec r$ and $t$ leads to,\\
\begin{equation}
\frac{\partial \rho}{\partial t} + \vec \nabla .
(\rho \vec v) = 0\label{e3}
\end{equation}
\begin{equation}
\frac{1}{\hbar} \frac{\partial S}{\partial t} + \frac{1}{2m} (\vec \nabla S)^2
+ \frac{V}{\hbar^2} - \frac{1}{2m} \frac{\nabla^2 R}{R} = 0\label{e4}
\end{equation}
where $\rho = R^2, \vec v = \frac{\hbar}{m} \vec \nabla S$
and $Q \equiv - \frac{\hbar^2}{2m} (\nabla^2 R/R)$.\\

Using the theory of fluid flow, it is well known that (\ref{e3}) and (\ref{e4})
lead to the Bohm alternative formulation of quantum
mechanics. In this theory there is a hidden variable namely the definite
value of position while the so called Bohm potential $Q$ can
be non local, two features which do not find favour with physicists.\\

Let us now consider the stationary solutions in the above formulation
viz. equation (\ref{e4}),
in the absence of external fields. As is known, the non local quantum
potential is given by
$$Q = \mbox{constant} = E,$$
the energy of the system. Further the velocity field is solenoidal,
\begin{equation}
\vec \nabla . \vec v = 0\label{e5}
\end{equation}
Remembering that S the phase is undefined up to a term which is a
multiple of $\pi$, we can now see from equation (\ref{e5}) that
there is a circulation which is given by (cf. ref.\cite{r1}).
\begin{equation}
\Gamma = \int_c \vec v . \vec dr = (\hbar/m) \int_c \vec \nabla S.d\vec r =
(\hbar/m)\oint dS = \frac{\pi \hbar n}{m}, n = 1,2,...\label{e6}
\end{equation}
For reasons which will become clear below we consider the ultra
relativistic case, $|\vec v | = c$ for all
particles of the fluid. We now get from equation (\ref{e6}),
\begin{equation}
m \Gamma = \int_c m \vec v . \vec dr = \frac{nh}{2},
n = 1,2,...\label{e7}
\end{equation}
where, an integration over all elements $\rho$, is implied. Here $n$ is
the number of nodes (or, in three dimensions, the end points of nodal
lines). We can immediately identify (\ref{e7}) with the quantum mechanical
spin $\frac{n}{2}$. Interestingly there are $2 \times \frac{n}{2} + 1 =
n+1$ multiply connected regions, both in hydro-dynamics and in the theory
of spin.\\

It is also worth noting that in (\ref{e7}), if the radius of the vortex
is taken to be $l$, then $l$ turns out to be the Compton wavelength, which
thus appears as a fundamental length. This will be commented upon later.\\

Further, considering for simplicity the vortex in (\ref{e6}) to be a thin
ring of radius $l$, we get
$$E = \oint \rho c^2 ds = mc^2,$$
where $\rho$ is the (line) density. Further from (\ref{e7}) we get
$$mc \oint ds = \frac{nh}{2},$$
whence, taking $n = 1,$
$$l = \frac{\hbar}{2mc}$$
The physical picture is now clear\cite{r3}. A particle can be pictured
as a fluid vortex which is steadily circulating along a ring (or in three
dimensions, a spherical shell) with radius equal to the Compton wavelength
and with velocity equal to that of light. Its total energy is given by,
as seen above
\begin{equation}
Q = E = mc^2\label{e8}
\end{equation}
and its angular momentum, which in quantum theory is quantized is given
by (\ref{e7}).
\section{The Origin of Inertial Mass}
We will now compare the above conclusions with the results of\cite{r4,r5}.
Our starting point is an equation deduced by Feynman\cite{r6} in a simple
way,
\begin{equation}
\imath \hbar \frac{\partial C (x)}{\partial t} = \frac{- \hbar^2}{2m'}
\frac{\partial^2 C(x)}{\partial x^2}\label{e9}
\end{equation}
where $C(x) \equiv | \psi (x) >$ is the probability amplitude
for the particle to be at the point $x$ at some given moment of time.\\
To deduce equation (\ref{e9}) we follow the development of\cite{r6} and
define a complete set of base states by the subscript $\imath \mbox{and}
U (t_2,t_1)$ the time elapse operator that denotes the passage of time between instants
$t_1$ and $t_2$, $t_2$ greater than $t_1$. We denote by, $C_\imath (t) \equiv
< \imath |\psi (t) >$, the amplitude for the state $|\psi (t) >$ to be in
the state $| \imath >$ at time $t,$ and
$$< \imath |U|j > \equiv U\imath j, U \imath j(t + \Delta t,t) \equiv
\delta_{\imath j} - \frac{\imath}{\hbar} H_{\imath j}(t)\Delta t.$$
We can now deduce from the super position of states principle that,
$$C_\imath (t + \Delta t) = \sum_{j} [\delta_{\imath j} - \frac{\imath}
{\hbar} H_{\imath j}(t)\Delta t]C_j (t)$$
and finally, in the limit,
\begin{equation}
\imath \hbar \frac{dC_\imath (t)}{dt} = \sum_{j} H_{\imath j}(t)C_j(t)\label{e10}
\end{equation}
where the matrix $H_{\imath j}(t)$ is identified with the Hamiltonian operator.
(To facilitate comparison we stick to the notation and development as given
in\cite{r6}. Before proceeding to derive the Schrodinger equation, we apply
equation (\ref{e10}) to the simple case of a two state system $(\imath, j =
1,2$ respectively; (cf.ref.\cite{r6}). This will provide a physical
picture for the later work. For a two state system we have
$$\imath \hbar \frac{dC_1}{dt} = H_{11}C_1 + H_{12}C_2$$
$$\imath \hbar \frac{dC_2}{dt} = H_{21}C_1 + H_{22}C_2$$
leading to two stationary states of energies $E-A$ and $E+A$, where
$E \equiv H_{11} = H_{22}, A = H_{12} = H_{21}.$ We can choose our zero
of energy such that $E = 2A.$ Indeed as has been pointed out by Feynman,
when this consideration is applied to the hydrogen molecular ion, the fact
that the electron has amplitudes $C_1$ and $C_2$ of being with either of
the hydrogen atoms, manifests itself as an attractive force which binds
the ion together, with an energy of the order of magnitude $A = H_{12}.$\\
To proceed, we consider in (\ref{e10}), the $\imath$ to be the space point
$x_\imath$ and we denote $C(x_n) \equiv C_n$ the probability amplitude for the
particle to be at this space point. Further let $x_{n+1} - x_n = b$.
Then considering only the point $x_n$ and its neighbours $x_{n\pm 1}$,
the equation (\ref{e10}) goes over into
\begin{equation}
\imath \hbar \frac{\partial C(x_n)}{\partial t} = EC(x_n) - AC(x_n - b) -
AC(x_n + b)\label{e11}
\end{equation}
In the limit $b \to 0$, with our choice of the arbitrary zero of energy,
(\ref{e11}) goes over into equation (\ref{e9}) where we have now dropped
the subscript distinguishing the space point, and $m' = \hbar^2/2Ab^2.$\\
We now observe that while equation (\ref{e9}) resembles the free
Schrodinger equation, as has been pointed out by Feynman, $m'$ is not really
the inertial mass, but an "effective mass" that emerges from the probability
amplitude for the particle to be found at a neighbouring point. So (\ref{e9})
is not the Schrodinger equation.\\
The Schrodinger equation can be obtained from (\ref{e9}) if it can be shown
that $m'$ can somehow be replaced by $m$. This is what we propose to do.\\
To start with let us suppose that the particle has no mass other than the
effective mass $m'$, so that we can treat equation (\ref{e9}) as the
Schrodinger type equation for such a particle which has only amplitude
to be at neighbouring points. Let us now suppose that the particle
acquires non zero probability amplitude to be present non locally at other
than neighbouring points. We can then no longer work with equations
(\ref{e11}) and (\ref{e9}). We will have to use the full equation (\ref{e10})
which explicitly exhibits this possibility. We rewrite equation (\ref{e10})
as
$$\imath \hbar \frac{dC_\imath (t)}{dt} = H_{\imath \imath}C_\imath(t) +
H_{\imath, \imath-1}C_{\imath -1}(t) +
H_{\imath,\imath+1}C_{\imath +1}(t)\\
+ \sum_{j} H_{\imath,\imath +j}(t)C_j(t), (j = \pm 2, \pm 3,)$$
or as in the transition of equation (\ref{e11}) to equation (\ref{e9}),
\begin{equation}
\imath \hbar \frac{\partial C(x)}{\partial t} = \frac{-\hbar^2}{2m'}
\frac{\partial^2 C(x)}{\partial x^2} + \int H(x,x')C(x')dx'\label{e12}
\end{equation}
where we have replaced $H_{\imath j}$ by $H(x,x')$ and the points $x_\imath$
are in the limit taken for the time being to be a continuum. This is as in the well known case of
the non-local Schrodinger equation
for a non-local potential\cite{r7} but for a particle having only an effective
mass.\\
The matrix $H(x,x')$ gives the probability amplitude for the particle at
$x$ to be found at $x'$, that is,
\begin{equation}
H(x,x') = < \psi (x')|\psi (x) >\label{e13}
\end{equation}
where as is usual we write $C(x)\equiv \psi (x)(\equiv | \psi (x) >$,
the state of a particle at the point $x$).\\
Usually the amplitude $H(x,x')$ is non-zero only for neighbouring points
$x$ and $x'$, that is, $H(x,x') = f(x)\delta (x-x').$ But if $H(x,x')$ is
not of this form, then there is a non-zero amplitude for the particle to
"jump" to an other than neighbouring point. In this case $H(x,x')$ may be
described as a non local amplitude. Indeed such non-local amplitudes are
implicit in the Dirac equation also and this will be commented on later.\\
We now give a quick derivation of how the inertial mass emerges from
equation (\ref{e12}). The non local Schrodinger equation (\ref{e12}), given
only the effective mass $m'$, can be written, with the help of (\ref{e13}),
as,
\begin{equation}
\imath \hbar \frac{\partial \psi}{\partial t} = \frac{-\hbar^2}{2m'}
\frac{\partial^2 \psi}{\partial x^2} + \int \psi^* (x')\psi (x) \psi (x')
U(x')dx',\label{e14}
\end{equation}
where,\\
i)$U(x) = 1$ for $|x| < R,R$ arbitrarily large and also $U(x)$ falls off
rapidly as $|x| \to \infty; U(x)$ has been introduced merely to ensure
the convergence of the integral; and\\
ii)$H(x,x') = < \psi (x')\psi (x) > = \psi^* (x')\psi (x).$\\
(\ref{e14}) is an integro-differential equation of degree three.\\
The presence of the, what at first sight may seem troublesome, non-linear
and non-local term, viz., the last term on the right side of (\ref{e14}) will
be satisfactorily explained in the sequel.\\
In (\ref{e14}),in the first approximation $\psi (x)$ can be taken to be the
solution of the Schrodinger like equation (\ref{e9}), viz.,
\begin{equation}
\imath \hbar \frac{\partial \psi}{\partial t} = \frac{-\hbar^2}{2m'}
\frac{\partial^2 \psi}{\partial x^2}\label{e15}
\end{equation}
In effect, we linearize (\ref{e14}), so that we get,
\begin{equation}
\imath \hbar \frac{\partial \psi}{\partial t} = [ - \frac{\hbar^2}{2m'}
\frac{\partial^2}{\partial x^2} + m_0]\psi \label{e16}
\end{equation}
where,
$$m_0 = \int \psi^* (x') \psi (x')U(x')dx'$$
In operator language, (\ref{e16}) becomes,
\begin{equation}
\bar H = \frac{p^2}{2m'} + m_0\label{e17}
\end{equation}
where $\bar H$ is the Hamiltonian operator, $p$ the momentum operator and where,
what can now be anticipated as a rest mass like term $m_0$, appears for a
particle assumed not to have any rest mass in the absence of the non-local
amplitude term in (\ref{e14}). Also we have replaced the Hamiltonian matrix
$H$ by $\bar H$ to stress that, to start with, in (\ref{e12}) and (\ref{e14}),
the particle has no inertial mass. To facilitate comparison with the usual
theory, we next multiply both sides of (\ref{e17}) by the constant
$\frac{m'}{m}$, where,
$$m = (m_0m')^{\frac{1}{2}}/c,$$
$c$ being the velocity of light. (The reason for the appearance of the
velocity of light, $c$ can be seen below (cf.equation (\ref{e19})) and the
constant could be absorbed into the state vector, whose direction is all
that matters. We then get,
\begin{equation}
\hat H = \frac{p^2}{2m} + mc^2\label{e18}
\end{equation}
The physical meaning of (\ref{e18}) is now clear. In an expansion of the
classical relativistic expression for energy,
$$E = (p^2c^2 + m^2c^4)^{1/2}$$
as is well known, if we keep terms up to the order $(p/mc)^2$, we get,
\begin{equation}
E = \frac{p^2}{2m} + mc^2\label{e19}
\end{equation}
We can now easily identify $m$ in (\ref{e18}) with the rest mass on
comparing this equation with (\ref{e19}). (Interestingly it is not
accidental that equation (\ref{e18}) corresponds to the approximation
(\ref{e19}) as will be seen below). If further, we denote
$$H = \hat H - mc^2,$$
where $H$ can be easily identified with the usual kinetic energy operator
(or energy operator in non-relativistic theory, remembering that we are
considering a free particle only), (\ref{e18}) becomes
\begin{equation}
H = \frac{p^2}{2m}\label{e20}
\end{equation}
In a strictly non relativistic context, where the rest energy of the
particle is not considered, the Hamiltonian is given by (\ref{e20});
otherwise, it is given approximately by (\ref{e18}). We get from
(\ref{e20}), the Schrodinger equation,
\begin{equation}
\imath \hbar \frac{\partial \psi}{\partial t} = - \frac{\hbar^2}{2m}
\frac{\partial^2 \psi}{\partial x^2}\label{e21}
\end{equation}
All these considerations can be considered in a postulative development
\cite{r5} and also generalized in a simple way to three dimensions,
but as there is no new physical insight, the details are not given.\\
The physical origin of the rest mass is clear from equation (\ref{e18}):
in the two state hydrogen molecular ion case considered earlier, it was the
amplitude for the single electron to be with one hydrogen atom or the
other which showed up as a binding energy. Similarly the amplitude of a
particle to be at $x$ or $x'$ viz. the second term on the right side of
equation (\ref{e14}) manifests itself as an (attractive) energy, which may be
called the mass energy of the particle or the self energy or the energy of
self interaction. This can be seen to be the particle's inertial mass.\\
We now come to the non local term in equation (\ref{e14}), the term which
gives the inertial mass. Non locality implies superluminal velocities
and the breakdown of causality which is not permissible in general.
However without any contradiction to the theory it is well known that
Quantum Mechanics allows such non locality, owing to the uncertainity
principle \cite{r8}, within the Compton wavelength of a particle.
So there is no contradiction if the non local integral in (\ref{e14})
is taken within the region of the particle's Compton wavelength, that is,
the inertial mass is a result of non local processes within the Compton
wavelength of the particle.\\
Indeed the usual Dirac equation also has a non local character: The operator
$c\vec \alpha . \vec p + \beta mc^2$ is equivalent to and replaces the
non-local square-root operator, $(- \hbar^2 \nabla^2 + m^2c^4)^{1/2}$. Here
also, the non-local effects in the form of negative energies are
encountered - again within the Compton wavelength region (cf.ref.\cite{r9}).\\
In the light of the preceding considerations, we can derive the Schrodinger
equation from an alternative angle: It appears that the "point" particle is
really spread over the non-locality region $\sim \bar b = \frac{\hbar}{mc}$, the
Compton wavelength. Further, the energy of the particle i.e., the energy
tied up within this region viz., 2A is the inertial mass energy $mc^2$. We
could now, speak of the amplitude for the particle at $x$ to be found
(locally) at a neighbouring point $x+b$, except that in the limit, $b \to
\bar b$ (and not as earlier 0). The effective mass $m'$ in equation (\ref{e9}) is then
given by,
$$m' = \frac{\hbar}{2Ab^2} = m,$$
that is the mass itself!\\
So, equation (\ref{e9}) can be interpreted as the Schrodinger equation.\\
It is worth re-emphasizing that it is the force of binding of non-local
positions within the Compton wavelength, rather like the Hydrogen molecular
ion binding, that manifests itself as inertial mass.\\
Finally we briefly comment on the appearance of the extra mass energy term
in equations like (\ref{e12}), (\ref{e14}), (\ref{e17}), (\ref{e18}) or
(\ref{e19})\cite{r5,r10,r11}.\\
 The Schrodinger equation is really the limiting case
of the Dirac equation in which process an inessential phase
factor is dropped. Another way of looking at this is that the constant
potential $m_o c^2$ does not affect the dynamics. That is the reason
why the Schrodinger equation is not Galilean invariant, as a non relativistic theory
should be, and infact exhibits the Sagnac effect, which a strictly
Galilean invariant theory should not\cite{r12}.\\

The convergence of the above formulation and the Bohm hydrodynamical
formulation is evident once we restrict ourselves to the
Compton wavelength and luminal velocities. The particle is now a
relativisitc fluid vortex circulating along a ring of radius equal to
the Compton wavelength. The Q given by (\ref{e8}) is the energy of
this system or particle and corresponds to the inertial mass term given
by the integral in (\ref{e14}) or equivalently the constant potential
term in (\ref{e19}).
\section{Monopoles}
It is well known that there is a close connection between the
hydrodynamic theory discussed in Section 2 and Dirac's theory of monopoles
\cite{r13}. The
starting point in this latter case is precisely the decomposition of the
wave function (\ref{e2}), but the focus is on the phase function $S$
which need not be integrable: exactly as in the case of the vortex above,
there can be nodal singularities. Infact the $S$ in this theory
defines the function $\vec K$ of Dirac, exactly as it does the momentum vector of
section 2. But this time $(\vec K, K_0)$ is identified with the
electromagnetic potential and an integral like (\ref{e7}) then consists
of, in addition to the term $\frac{n \hbar}{2}$ the electromagnetic flux,
again $n$ being the number of nodal lines with end points inside the
vortex or region of integration. Thus the well known equation of the
magnetic monopole viz. $\mu = \frac{1}{2} n \hbar \frac{c}{e},$ on
identifying
\begin{equation}
\vec K \equiv \frac{e}{\hbar c} \vec H\label{e22}
\end{equation}
($\vec H$ being the magnetic field) with the momentum of section 2 gives
back equation (\ref{e7}) for quantized spin. We will come back to this
point later.
\section{Zitterbewegung and the Compton Wavelength}
We will now examine briefly the phenomenon of zitterbewegung in the context
of the Compton wavelength. In the usual theory of the Dirac equation
\cite{r14}, it is well known that
the eigen values of the velocity operator $c \vec \alpha \mbox{are} \pm c,$
the velocity of light while the position operator is non Hermitian: It consists
of a real part which is the usual position and a rapidly oscillating (or
zitterbewegung) imaginary part,
\begin{equation}
x = (c^2 p_1 H^{-1}t) + \frac{\imath}{2} c \hbar (\alpha_1 - cp_1 H^{-1})
H^{-1}\label{e23}
\end{equation}
Both these puzzling facts are reconciled by the fact that our measurements
are really averaged over time intervals of the order $\hbar/mc^2$
and correspondingly over the space intervals of the order of $\hbar/mc$,
the Compton wavelength. In
this case the imaginary part in (\ref{e23}) disappears (cf. ref.\cite{r14}).
Hermiticity and Physics begins after such an averaging necessitated by
our gross measurements.\\

One could say that (\ref{e23}) applies in a non local region bounded by the
Compton wavelength as we saw in section 2. Within the region, we have to
contend with unphysical phenomena like superluminal velocities and negative
energies and in general non Hermitian operators. Outside the Compton
wavelength, that is on averaging over space time intervals of this order,
we are back in usual Physics.\\

We consider now, for simplicity, the free particle Dirac equation. The
solutions are of the type,
\begin{equation}
\psi = \psi_A + \psi_S\label{e24}
\end{equation}
where
$$
\psi_A =   e^{\frac{\imath}{\hbar} Et} \ \left(\begin{array}{l}
                                          0 \\ 0 \\ 1 \\ 0
                             \end{array}\right) \mbox{ or } \ e^{\frac{\imath}{\hbar} Et}
                          \ \   \left(\begin{array}{l}
                                 0 \\ 0 \\ 0 \\ 1
                              \end{array}\right) \mbox{ and }
$$
\begin{equation}
\label{e25}
\end{equation}
$$
\psi_S =  e^{-\frac{\imath}{\hbar} Et} \ \left(\begin{array}{l}
                                 1 \\ 0 \\ 0 \\ 0
                               \end{array}\right) \mbox{ or } e^{-\frac{\imath}{\hbar} Et}
                            \ \   \left(\begin{array}{l}
                                 0 \\ 1 \\ 0 \\ 0
                               \end{array}\right)
$$
denote respectively the negative energy and positive energy solutions. From
(\ref{e24}) the probability of finding the particle in a small volume
about a given point is given by
\begin{equation}
| \psi_A + \psi_S|^2 = |\psi_A|^2 + |\psi_S|^2 +
(\psi_A \psi_S^* + \psi_S \psi_A^*)\label{e26}
\end{equation}
Equations (\ref{e25}) and (\ref{e26}) show that the negative energy and
positive energy solutions form a coherent Hilbert space and so the
possibility of transition to negative energy states exists. This difficulty
however is overcome by the Hole theory which uses the Pauli exclusion
principle.\\

However the last term on the right side of (\ref{e26}) is like the
zitterbewegung term. When we remember that we really have to consider
averages over space time intervals of the order of $\hbar/mc$ and
$\hbar/mc^2$, this term disappears and effectively the negative energy
solutions and positive energy solutions stand decoupled in what is now
the physical universe.\\

A more precise way of looking at this is\cite{r15} that as is well known, for
the homogeneous Lorentz group, $\frac{p_0}{|p_0|}$ commutes with all operators
and yet it is not a multiple of the identity as one would expect according
to Schur's lemma: The operator has the eigen values $\pm 1$ corresponding
to positive and negative energy solutions. This is a super selection
principle pointing to the two incoherent Hilbert spaces or universes
\cite{r16} now represented by states $\psi_A$ and $\psi_S$ which have
been decoupled owing to the averaging over the Compton wavelength space-
time intervals. But all this refers to energies such that our length scale
is greater than the Compton wavelength. As we reach energies corresponding
to the Compton wavelength scale, negative energy solutions show up as
anti particles. Thus the super selection principle which comes into
play on averaging over Compton wavelength scales dispenses with the
Pauli exclusion principle.\\
Thus once again we see that outside the Compton wavelength region we
recover the usual physics.
\section{A Classical Viewpoint}
Let us now try to understand from the standpoint of classical theory, why
we encounter luminal (or superluminal) velocities and complex coordinates,
corresponding to non-Hermitian operators, within the Compton wavelength
region. From a classical point of view, we could say that if in the Lorentz
transformation,
\begin{equation}
x = \gamma (x'- v t), \gamma = (1 - v^2/c^2)^{-1/2}\label{e27}
\end{equation}

$v > c$ is allowed, then the coordinates become imaginary, this being
true within the Compton wavelength as in (\ref{e23}), in the sense that non locality is
allowed there. So (\ref{e23}) can be understood as representing a
coordinate which is imaginary within the Compton wavelength but becomes
the usual position coordinate outside, that is after averaging over
these intervals. One way of interpreting (\ref{e23}) would be that from
our physical point of view using (\ref{e27}) there is a region where $v > c$,
consisting of virtual or superluminal ghost particles bounded by a region,
a sphere of radius equal to the Compton wavelength consisting of massless
"particlets" (to distinguish them from partons, instantons and the like, or to make
a clean break, "Ganeshas") with velocity of light. Only on averaging
over this vortex like sphere or region, do we come to the domain of
conventional physics and the usual particles moving with sub luminal
velocities. It may be remarked that the De Broglie-Bohm picture of a
particle is that of an average over an ensemble (cf. ref.\cite{r1}) but
the above picture is different: It is an averaging over a physically
inaccessible region.\\

Indeed it is known that for a collection of relativistic particles, the
various mass centres form a two-dimensional disc perpendicular to the
angular momentum vector $\vec L$ and with radius (ref.\cite{r17})
\begin{equation}
r = \frac{L}{mc}\label{e28}
\end{equation}

Further if the system has positive energies, then it must have an
extension greater than $r$, while at distances of the order of
$r$ we begin to encounter negative energies.\\

If we consider the system to be a particle of spin or angular momentum
$\frac{\hbar}{2}$, then equation (\ref{e28}) gives, $r = \frac{\hbar}{2mc}$.
That is we get back the Compton wavelength.\\

On the other hand it is known that (cf. ref.\cite{r9}), if a Dirac particle
is represented by a Gausssian packet, then we begin to encounter negative
energies precisely at the same  Compton wavelength as above. Thus a
particle can indeed be treated as a vortex or a spherical shell of
relativistic sub constituents or "particlets" (or Ganeshas).\\
Another way of looking at this from a hydrodynamical perspective is to
consider for example a one dimensional streamlined flow of a fluid, with
velocity $\vec v \equiv v_x$, along the $x$ axis. In this case
$\vec \nabla \times \vec v$ vanishes everywhere, that is there is no
circulation (no vortices).\\
However all this applies to a singly connected space. As
is well known one could still have a circulation about a point $P$ (for
example cf.ref.\cite{r18}) exactly as in the case of equation (\ref{e6})
because $P$ would be a singularity and the space would no longer be
singly connected, rather it would be doubly connected. In this case
$\vec \nabla \times \vec v$ would vanish everywhere except on the
boundary of a closed curve around the point $P$. We would now have a
complex vector potential in the complex plane $x+\imath y$ (cf.ref.\cite
{r18}). Away from the point $P$ in what may be called the asymptotic
region we would have the one dimensional flow, but as we approach the
point $P$, that is the boundary of the curve enclosing $P$, we
encounter circulatory motion in the $x+\imath y$ plane.\\
In our case the region bounded by the Compton wavelength plays the
role of the closed curve around the point $P$. Outside this region we
have the usual space (or space time) of Physics. But as we approach the
Compton wavelength region we encounter a region where each
of the space time axes becomes as it were a complex plane. We will
return to this point.\\
\section{Particles as Black Holes}
The fact that, as we saw in sections 2 and 3, the mass generating
non-local amplitudes are confined to a
region of width $\sim \frac{\hbar}{mc}$ suggests that the particle could
be a black hole, because in this case also, there is a width, the horizon,
inside which such unphysical phenomena appear. The possibility that a
particle could be a Schwarzchild black hole has been examined earlier by
Markov, Motz and others\cite{r19,r20,r21,r22} and leads to a high particle mass of
$10^{-5}gm$, without much insight into other properties.\\
So let us
approach the problem from a different angle. We consider a charged Dirac
(spin half) particle. If we treat this as a spinning black hole, there is an
immediate problem:The horizon of the Kerr-Newman black hole becomes in
this case, complex\cite{r23},
\begin{equation}
r_+ = \frac{GM}{c^2} + \imath b_,b \equiv (\frac{G^2 Q^2}{c^8} + a^2 -
\frac{G^2M^2}{c^4})^{1/2}\label{e29}
\end{equation}
where $G$ is the gravitational constant, $M$ the mass and $a \equiv
L/Mc,L$ being the angular momentum. That is, we have a naked singularity
apparently contradicting the cosmic censorship conjecture. However, in the
Quantum Mechanical domain, (\ref{e29}) can be seen to be meaningful.\\
Infact, the position coordinate for a Dirac particle as we have seen is
given by Dirac\cite{r14}
\begin{equation}
x = (c^2p_1 H^{-1}t + a_1) + \frac{\imath}{2}c\hbar
(\alpha_1 - cp_1 H^{-1})H^{-1},\label{e30}
\end{equation}
where $a_1$ is an arbitrary constant and $c\alpha_1$ is the velocity
operator with eigen values $\pm c$. The real part in (\ref{e30}) is the
usual position while the imaginary part arises from zitterbewegung.
Interestingly, in both (\ref{e29}) and (\ref{e30}), the imaginary part is
of the order of $\frac{\hbar}{mc}$, the Compton wavelength, and leads to
an immediate identification of these two equations. We must remember that
our physical measurements are gross as noted earlier - they are really
measurements averaged over a width of the order $\frac{\hbar}{mc}$.
Similarly, time measurements are imprecise to the tune $\sim \frac{\hbar}
{mc^2}$. Very precise measurements if possible, would imply that all
Dirac particles would have the velocity of light, or in the Quantum
Field Theory atleast of Fermions, would lead to divergences. (This is
closely related to the non-Hermiticity of position operators in
relativistic theory as can be seen from equation (\ref{e30}) itself
\cite{r15}. Physics as pointed out earlier begins after an averaging over the above
unphysical space-time intervals. In the process as is known (cf.ref.
\cite{r15}), the imaginary or non-Hermitian part of the position operator
in (\ref{e30}) disappears. That is in the case of the QMBH (Quantum
Mechanical Black Hole), obtained by identifying (\ref{e29}) and
(\ref{e30}), the naked singularity is shielded by a Quantum Mechanical
censor.\\
To continue, we first adhoc treat a Dirac particle as a Kerr-Newman
black hole of mass $m$, charge $e$ and spin $\frac{\hbar}{2}$. The
gravitational and electromagnetic fields at a distance are given by
(cf.ref.\cite{r24},
\begin{eqnarray}
\Phi (r) = - \frac{Gm}{r} + 0 (\frac{1}{r^3}) E_{\hat r} = \frac{e}{r^2} +
0 (\frac{1}{r^3}),E_{\hat \theta} =
0 (\frac{1}{r^4}),E_{\hat \phi} = 0, \nonumber \\
B_{\hat r} = \frac{2ea}{r^3} cos \theta + 0 (\frac{1}{r^4}),
B_{\hat \theta} = \frac{easin\theta}{r^3} + 0(\frac{1}{r^4}),
B_{\hat \phi} = 0,\label{e31}
\end{eqnarray}
exactly as required. Infact, as is well known, (\ref{e31}) also exhibits the
electron's anomalous gyromagnetic ratio $g = 2.$ So we are on the right
track!\\
We next examine more closely, this identification of a Dirac particle with
a Kerr-Newman black hole. We reverse the arguments after equation (\ref{e30})
which lead from the complex or non-Hermitian coordinate operators to Hermitian
ones: We consider instead the displacement,
\begin{equation}
x^\mu \to x^\mu + \imath a^\mu\label{e32}
\end{equation}
and first consider the temporal part, $t \to t + \imath a^0$, where
$a^0 \approx \frac{\hbar}{2mc^2}$, as before. That is, we probe into the
QMBH or the zitterbewegung region inside the Compton wavelength as suggested
by (\ref{e29}) and (\ref{e30}). Remembering that $|a^\mu | < < 1$, we have,
for the wave function,
$$\psi (t) \to \psi (t+\imath a^0) = \frac{a^0}{\hbar} [\imath \hbar
\frac{\partial}{\partial t} + \frac{\hbar}{a^0}]\psi (t)$$
As $\imath \hbar \frac{\partial}{\partial t} \equiv p^0$, the usual fourth
component of the energy momentum operator, we identify, by comparison with
the well known electromagnetism-momentum coupling, $p^0 - e\phi$, the
usual electrostatic charge as,
\begin{equation}
\Phi e = \frac{\hbar}{a^0} = mc^2\label{e33}
\end{equation}
In the case of the electon, we can verify that the equality (\ref{e33})
is satisfied:\\
We follow the classical picture of a particle as a rotating shell with
velocity $c$, as encountered in sections 2 and 6, and which will be
further justified in the sequel. The electrostatic potential inside a
spherical shell of radius $'a'$ is,
\begin{equation}
\Phi = \frac{e}{a}\label{e34}
\end{equation}
As is well known, the balance of the centrifugal and Coulomb forces gives,
for an electron orbiting another at the distance $a$,
$$a = \frac{e^2}{mc^2},$$
which is the classical electron radius.\\
So, (\ref{e34}) now gives,\\
$$e\Phi = mc^2,$$
which is (\ref{e33}).\\
If we now use the usual value of $'a'$ viz., $2.8 \times 10^{-13} cm.,$ in
(\ref{e34}) and substitute in (\ref{e33}), while rewriting the right side
as $\hbar c/(\hbar/mc)$ and substitute the value of the electron Compton
wavelength, $\frac{\hbar}{mc} = 3.8 \times 10^{-11} cm.,$ we get
$$\hbar c \approx 136 e^2$$
That is, we get the rationale for this fundamental relation, which no longer
turns out to be accidental. In any case, equation (\ref{e33}) throws up the
connection between the charge, mass and the velocity of light.\\
It may be noted in passing that in the usual displacement operator theory
(\cite{r14}) the operators like $\frac{d}{dx}$ or $\frac{d}{dt}$
are indeterminate to the extent of a purely imaginary additive constant
which is adjusted against the Hermiticity of the operators concerned.\\
We next consider the spatial part of (\ref{e32}), viz.,
$$\vec x \to \vec x + \imath \vec a, \mbox{where} |\vec a | = \frac{\hbar}
{2mc},$$
given the fact that the particle is now seen to have the charge $e$ (and mass
$m$). As is well known\cite{r25}, this leads in General Relativity
from the static Kerr metric to the Kerr-Newman metric where the gravitational
and electromagnetic field of the particle is given by (\ref{e31}), including
the anomalous factor $g = 2$. In General Relativity, the complex transformation
(\ref{e32}) and the subsequent emergence of the Kerr-Newman metric has no
clear explanation. Nor the fact that, as noted by Newman\cite{r26} spin is
the orbital angular momentum with an imaginary shift of origin. But in the
Quantum Mechanical context and in view of the considerations of section 2 and 6,
we can see the rationale: the origin of (\ref{e32}) lies in the QMBH. We
started with a massless particle. Then we saw the emergence of mass and also
the origin of gravitation and electromagnetism in the processes inside the
Compton wavelength represented by an imaginary displacement - the nonlocal
QMBH region.\\
More specifically, the temporal part of the transformation (\ref{e32}) lead
to the appearance of charge in (\ref{e33}). The space part then, as is known
leads to the Kerr-Newman metric.\\
There is another way to see the emergence of electromagnetism. It is well known
that for the Dirac four spinor,$(\theta_{\chi})$ where $\theta$ denotes the
positive energy two spinor and $\chi$ the negative energy two spinor, at
and within the Compton wavelength, it is $\chi$ that dominates. Further,
under reflections, while $\theta \to \theta, \chi$ behaves like a psuedo-
spinor\cite{r9}
$$\chi \to - \chi$$
Hence the operator $\frac{\partial}{\partial x^\mu}$ acting on $\chi$, a
density of weight $N = 1,$ has the following behaviour\cite{r27},
\begin{equation}
\frac{\partial \chi}{\partial x^\mu} \to \frac{1}{\hbar} [\hbar
\frac{\partial}{\partial x^\mu} - NA^\mu ]\chi\label{e35}
\end{equation}
where,
\begin{equation}
A^\mu = \hbar \Gamma^{\mu \sigma}_{\sigma} = \hbar \frac{\partial}
{\partial x^\mu} log (\sqrt{|g|}) \equiv \nabla^\mu \Omega\label{e36}
\end{equation}
As before we can identify $NA^\mu$ in (\ref{e35}) with the electro-magnetic
four potential. That $N = 1$, explains the fact that charge is discrete. It
will be shown in the next section that,
\begin{equation}
A^\mu \sim const. \frac{e^2}{r}\label{e37}
\end{equation}
in agreement with (\ref{e33}). That is, electromagnetism is the result of
the covariant derivative that arises due to the Quantum Mechanical behaviour
of the negative energy components within the Compton wavelength region.\\
We observe, that in case the mass $m \to 0$, the considerations of
section 3 imply that there are no negative energy components while (\ref{e33})
and (\ref{e37}) show that such a particle has no charge. The massless
neutrino fits this description exactly: it has a two component wave function
and is chargeless.\\
There is also the muon which satisfies (\ref{e37}) in the order of magnitude
sense. But it is unstable and disintegrates into an electron (or positron)
and two neutrinos anyway.\\
Thus these considerations describe the stable leptons, viz., electrons and
neutrinos, and approximately the remaining unstable lepton.\\
It is worth noting that equation (\ref{e36}) strongly resembles Weyl's
formulation for the unification of electromagnetism and gravity. But there
is an important difference \cite{r28}: Weyl's Christoffel symbol
contains two independant entities - the metric tensor $g^{\mu v}$ and the
electromagnetic potential $\Phi$. So there is no unification of electromagnetism
and gravity. Our formulation uses only the Quantum Mechanical pseudo spinor
property.\\
It is interesting that, in the light of the above considerations an application
of Maxwell's equations in this Compton wavelength region of "charged matter"
leads to meaningful results: In this case the fact that $A^\mu$ in (\ref{e36})
is a four gradient poses no problem. We have, using Maxwell's equations
\cite{r29},
\begin{eqnarray}
\phi \equiv A^0 = \frac{\partial \Omega}{\partial t}, \vec B = \vec \nabla
X \vec A = \vec \nabla X (\vec \nabla \Omega) = 0, \nonumber \\
\vec E = - \frac{\partial \vec A}{\partial t} - \vec \nabla \phi =
-2 \vec \nabla \Omega, \vec E = -2 \vec \nabla \phi\label{e38}
\end{eqnarray}
Also,
\begin{equation}
\vec \nabla. \vec E = 2\nabla^2 \phi = 4 \pi \sigma,
\vec \nabla. \vec B = 0,\label{e39}
\end{equation}
while
\begin{equation}
\vec \nabla \times \vec B = 0 = 4 \pi s + \frac{\partial \vec E}
{\partial t}\label{e40}
\end{equation}
and,
$$\vec \nabla \times \vec E = - \sigma \frac{\partial \vec B}{\partial t}
= \vec \nabla \times (2\vec \nabla \phi) = 0$$
Equations (\ref{e38}), (\ref{e39}), and (\ref{e40}) show that effectively
this is a steady field with potential $\phi$ that is, we would work as if
we have a steady field of potential $\phi$ except that there is an
anamolous doubling of the charge and current. Now, as is well known the
usual orbital magnetic moment is given by \cite{r30}
\begin{equation}
\mu = \frac{e}{2mc}p_\phi\label{e41}
\end{equation}
where $p_\phi$ is the angular momentum and $e$ is the charge. In our
case, $e$ in (\ref{e41}) is effectively replaced by $2e$, so that in the
usual units of $e/2mc$, we now have for the Dirac particle, instead of
(\ref{e41}),
$$g = \frac{\mu}{p_\phi} = 2$$
This is the anomalous gyromagnetic ratio which arises because as noted
earlier spin is the orbital angular momentum with an imaginary shift of
origin, or equivalently within the Compton wavelength region. We come to this
point now, in greater detail.
\section{A General Relativistic Approach: Origin of QCD Interactions}
Thus far it appears that the QMBH description applies to electrons and
more generally Leptons. In the light of the preceding considerations, we will
now approach the problem from a General Relativistic point of view. This will
also reveal the origin of QCD type interactions. Taking the cue from the
foregoing considerations, we now treat the particle as a relativistic fluid
of "particlets" (or Ganeshas). Our starting point is the linearized theory
\cite{r24}:
\begin{equation}
g_{\mu v} = \eta_{\mu v} + h_{\mu v}, h_{\mu v} = \int
\frac{4T_{\mu v}(t-|\vec x - \vec x'|,\vec x')}{|\vec x - \vec x'|}
d^3x'\label{e42}
\end{equation}
(A bar on T has been dropped.)\\
In (\ref{e42}), velocities comparable to the velocity of light $c$ are allowed
and also the stresses $T^{jk}$ and momentum densities $T^{0j}$ can be
comparable to the energy momentum density $T^{00}$. As in ref.\cite{r24}, we
can easily deduce that, when $\frac{|\vec x'|}{r} < < 1$, where
$r \equiv |\vec x |$, and in a frame with origin at the centre of mass and at
rest with respect to the particle,
\begin{equation}
Gm = \int T^{00} d^3 x\label{e43}
\end{equation}
\begin{equation}
S_k = \int \epsilon_{klm} x^{l}T^{m0}d^3 x\label{e44}
\end{equation}
where $m$ is the mass (or approximate mass because of the linear
approximation), and $S_k$ is the angular momentum. We next observe that,
\begin{equation}
T^{\mu v} = \rho u^\mu u^v\label{e45}
\end{equation}
If we now work in the Compton wavelength region  of the QMBH, we have,
while $u^0 = 1$,
\begin{equation}
|u^l| = c\label{e46}
\end{equation}
(This is the Quantum Mechanical input)\\
Substitution of (\ref{e45}) and (\ref{e46}) in (\ref{e44}) gives on using the
Mean Value Theorem,
$$S_k = c < x^l > \int \rho d^3 x$$
As $< x^l > \sim \frac{\hbar}{2mc}$, using (\ref{e43}), we get, $S_k \approx
\frac{\hbar}{2}$, as required for a spin half particle. Infact this relation
becomes exact if we treat the QMBH as effectively a rotating shell distribution
of radius $\hbar/2mc$ as noticed earlier and, keeping in mind the fact that
the interior region is in any case unphysical as seen in section 6, and
is described by complex space-time coordinates. Once again we can see why
orbital angular momentum with a complex shift gives spin, as noticed earlier
by Newman but without a rationale (cf.ref.\cite{r26}).\\
The gravitational potential can similarly be obtained from (\ref{e42}) and
(\ref{e43}) (cf. ref.\cite{r24}),
\begin{equation}
\Phi = - \frac{1}{2}(g^{00} - \eta^{00}) = - \frac{Gm}{r} + 0
(\frac{1}{r^3})\label{e47}
\end{equation}
We saw in section 7 that the electromagnetic potential is given by,
$$A^\mu = \hbar \Gamma^{\mu \sigma}_{\sigma}$$
Using the expression for the Christoffel symbols, we have,
$$A_\sigma = \frac{1}{2} (\eta^{\mu v} \hbar_{\mu v}), \sigma,$$
so that, from (\ref{e42}),
$$A_0 = 2 \int \eta^{\mu v} \frac{\partial}{\partial t}
[\frac{T_{\mu v}(t-|\vec x - \vec x'|, \vec x')}{|\vec x - \vec x'|}] d^3 x'$$
Remembering that $| \vec x - \vec x'| \approx r$ for the distant region
we are considering, we have,
$$A_0 \approx \frac{2}{r} \int \eta^{\mu v} [\frac{\partial}{\partial \tau}
T_{\mu v}(\tau,\vec x'). \frac{d}{dt}(t-|\vec x - \vec x'|)]
d^3 x' \approx \frac{2}{r} \int \eta^{\mu v} \frac{d}{d\tau} T_{\mu v}.
(1+c)d^3x',$$
or finally
\begin{equation}
A_0 \approx \frac{2c}{r} \int \eta^{\mu v} \frac{d}{d\tau} T_{\mu v} d^3 x'\label{e48}
\end{equation}
as $c > > 1$, and where we have used the fact that in the Compton wavelength
region, $|u_v| = c$.\\
It has already been observed that QMBH can be treated as a rotating shell
distribution with radius $R \equiv \frac{\hbar}{2mc}.$ So we have,
\begin{equation}
|\frac{du_v}{dt}| = |u_v|\omega\label{e49}
\end{equation}
where $\omega$, the angular velocity is given by,
\begin{equation}
\omega = \frac{|u_v|}{R} = \frac{2mc^2}{\hbar}\label{e50}
\end{equation}
We get the same relation in the theory of the Dirac equation, remembering
that in (\ref{e43}) and (\ref{e44}) the centre of mass is at rest:
$$\imath \hbar \frac{d}{dt} (c\alpha_\imath) = -2mc^2 (c\alpha_\imath),$$
where $c\alpha_\imath$ is the velocity operator (cf.ref.\cite{r14}). Finally,
on using (\ref{e45}), (\ref{e49}) and (\ref{e50}) in (\ref{e48}), we get,
\begin{equation}
\frac{e'e}{r} = A_0 \sim \frac{\hbar c^3}{r} \int \rho \omega d^3 x'
\sim (Gmc^3) \frac{mc^2}{r}\label{e51}
\end{equation}
where $e' = 1 esu$ corresponds to the charge $N = 1$ and $e$ is the test
charge.\\
Because of the approximations taken in deducing (\ref{e51}), a dimensional
constant $(\frac{L}{T})^5$ has to be multiplied on the left side, which then
becomes, in units, $c = G = 1,$
$$e'e.(\mbox{dimensional  constant}) \approx 1.6 \times 10^{-111} cm^2$$
The right side is,
$$Gm^2 c^5 \approx 4.5 \times 10^{-111} cm^2,$$
in broad agreement with the left side.\\
Alternatively, using the values of $G,m$ and $c$ in (\ref{e51}), we get,
$$e \sim 10^{-10} esu,$$
which is correct.\\
Yet another way of looking at (\ref{e51}) is, that we get, as $e' = 1
esu \sim 10^{10}$
$$\frac{e^2}{Gm^2} \sim 10^{40},$$
which is well known empirically. But equation (\ref{e51}) gives the reason
for this relation.\\
In any case, equations (\ref{e33}) and (\ref{e51}) show the inter-relation
between $e,m,c \mbox{and} G.$\\
So far we have been considering distances far from the particle: $|\vec x' -
\vec x| > > |\vec x'|.$ This is the approximation invoked in a transition
from (\ref{e42}) to equations (\ref{e43}), (\ref{e44}) etc. Let us now see
what happens when $|\vec x| \sim |\vec x'|.$ In this case, we have from
(\ref{e42}), expanding in a Taylor series about $t$,
\begin{eqnarray}
h_{\mu v} = 4 \int \frac{T_{\mu v}(t,\vec x')}{|\vec x - \vec x'|}d^3 x'+
(\mbox{terms   independent  of}\vec x) + 2 \nonumber \\
\int \frac{d^2}{dt^2} T_{\mu v} (t,\vec x'). |\vec x - \vec x'| d^3 x' +
0(|\vec x - \vec x'|^2)\label{e52}
\end{eqnarray}
The first term gives a Coulombic $\frac{\alpha}{r}$ type interaction except
that the coefficient $\alpha$ is of much greater magnitude as compared to
the gravitational or electromagnetic case, because in this approximation,
in an expansion of $(1/|\vec x - \vec x'|),$ all terms are of comparable order.
To proceed further, using (\ref{e49}), we have,
$$\frac{d}{dt} T^{\mu v} = \rho u^v \frac{du^\mu}{dt} + \rho u^\mu
\frac{du^v}{dt} = 2 \rho u^\mu u^v \omega,$$
so that,
$$\frac{d^2}{dt^2} T^{\mu v} = 4 \rho u^\mu u^v \omega^2 = 4 \omega^2 T^{\mu v}$$
where $\omega$ is given by (\ref{e50}). Substitution in (\ref{e52}) gives,
\begin{equation}
h_{\mu v} \approx - \frac{\beta M}{r} + 8\beta M
(\frac{Mc^2}{\hbar})^2.r\label{e53}
\end{equation}
$\beta$ being a constant.\\
This resembles the QCD quark potential \cite{r31}, with both the Coulombic
and confining parts. Taking for $M$ the mass of a typical $C$ quark $\sim
1.8 Gev$ (cf.ref.\cite{r31}), the ratio of the coefficients of the $r$
term and the $\frac{1}{r}$ term as obtained from (\ref{e53}) is $\sim
\frac{1}{\hbar^2}(Gev)^2$ as in the case of QCD (ref.\cite{r31}). In any case
these considerations show that we can get different interactions at
different distances in a unified picture, which can approximately atleast
represent quarks also.\\
In this picture, how do we accommodate anti-particles, for example positrons?
While treating the negative energy spinor as a density in section 7 we had
assumed that $N = 1$. Equally well, we could have chosen $N = -1$. This
reverses the sign of the charge, all else remaining the same. So with
$N = -1$, we get a positron. Similarly for quarks, $N$ can be taken to be
fractional. But this apart it must be remembered that whereas for electrons
we took the asymptotic expansions of equations like (\ref{e42}), in the case
of quarks we had to consider the region near the Compton wavelength itself.\\
Thus, it appears that the treatment of Leptons and approximately quarks as
QMBH leads to meaningful results. On the other hand, these are the most
fundamental constituents of matter, according to current thinking. An
alternative is suggested in the next section.
\section{Discussion and Miscellaneous Comments}
1) We have seen that a particle could be treated as a relativistic vortex, that is a vortex
where the velocity of circulation equals that of light or a spherical
shell, whose constituents are again rotating with the velocity of light
or as a black hole described by the Kerr-Newman metric for a spin $\frac{1}{2}$
particle.\\
The fact that we get the gravitational potential $\frac{m}{r}$ in equation
(\ref{e47}) again confirms that mass comes from the Compton wavelength region.\\
2) The equation (\ref{e6}) emerges on using the fact that $S$ is defined
only up to a multiple of $\pi$, whence we get equation (\ref{e7}) giving
quantized spin. As pointed out from equation (\ref{e7}) the Compton
wavelength emerges. On the other hand equation (\ref{e28}) shows that
given the spin $\frac{\hbar}{2}$, we get the Compton wavelength. It is
also to be noted that equation (\ref{e44}) gives the spin $\frac{\hbar}{2}$
if we use the Compton wavelength. The Compton wavelength itself appears in
quantum mechanics due to the Heisenberg uncertainity principle. So it
appears that the quantum mechanical quantized spin and Compton wavelength
can be obtained from classical considerations like relativistic vortices.
In any case the remarkable universality of the Compton
wavelengh was pointed out by Wigner\cite{r32} - the above considerations show
why it emerges in a natural way. It is interesting to note that Wignall\cite{r33}
has pointed out that it is the Compton wavelength which is primary, from
which even the mass follows. The foregoing lends support to this viewpoint.\\
It may be pointed out, that interestingly, from a different viewpoint, using
considerations of self symmetry, if we assume that the scale of the
universe is broken at some stage, that is there is an ultimate micro-level for
our measurements, then the Compton wavelength again appears as a fundamental
length \cite{r34}.\\
3) The fact that the spin of the particle is directly connected to the
number of end points of the nodal lines, as seen in section 2 appears to
indicate that Fermions are primary and that Bosons can be treated as
bound states of Fermions. As pointed out, quarks also
could be approximately treated as Quantum Mechanical Black Holes in the foregoing
sense, and as it is known pions and other hadrons are indeed treated as bound states of a
quark and an anti quark. (Indeed from considerations of the symmetry
between leptonic and hadronic currents, leptons and hadrons appear to
be the same \cite{r35}.)\\
4) In ordinary Quantum Mechanics, $\psi$ being the wave function, $\psi \psi^*$
is proportional to the \underline{probability} density. On the other hand,
we saw in section 3 that the mass density is produced by the non-linear
amplitude $\psi \psi^*$ in the Compton wavelength region. More specifically
we saw in sections 7 and 8 that it is $\chi$, the negative energy part of
the Dirac four spinor (which dominates in this region), that is relevant.
That is, $\rho$ being the \underline{material} density,
\begin{equation}
\rho \alpha \chi \chi^*\label{e54}
\end{equation}
As observed in section 7, for the two component neutrino, $\chi = 0,$ and
the neutrinos are massless.\\
It was shown in an earlier communication \cite{r36,r37}, how Gravitation can emerge from the Schrodinger equation
self-consistently. Again, it is the identification of the material density
in (\ref{e54}) which gives substance to that result.\\
5) Treating the particle as a vortex as in section 2, arguments for a monopole
in section 4 then show that there would be the Bohm-Ahranov like effect
\cite{r38} at the Compton wavelength scale.\\
6) If in the position formula (\ref{e30}), we consider the real part and
also a time interval of $\sim \frac{\hbar}{mc^2}$, we get for $p = mc,$
$$x = \frac{\hbar}{mc}$$
This result can also follow from the Heisenberg Uncertainity Principle.\\
We see the emergence of the
Planck constant $\hbar$ in the extreme situation of the maximum velocity
and minimum physical space-time intervals. Moreover, $h,m$ and $c$ are
inter-related. Taking the cue from here, we pick up the result in section
3 viz.,
$$E\alpha m \quad \mbox{or}\quad E = my,$$
where $y$ is the constant of proportionality which was identified adhoc
earlier with $c^2$. Using now Heisenberg's Uncertainity relations, and
considering the extreme case, we have, firstly,
$$t \sim \frac{\hbar}{my},$$
so that $x = ct \sim \frac{\hbar c}{my}$, where $c$ is the maximum possible
velocity. Further, in this case,
$$p = mc \sim \frac{\hbar}{x} = \frac{my}{c},$$
so that $y = c^2$.\\
This provides a Quantum Mechanical justification for the formula $E = mc^2$,
without taking recourse to special relativity. Indeed as suggested by
section 3, e.g. equation (\ref{e18}) and the discussion of the Sagnac effect,
the origin of special relativity could be traced to these Quantum
Mechanical considerations.\\
7) It is interesting to note that the above model of a particle
could give a rationale for the left handedness of the neutrino in the light of
sections 5 and 6. In the case of the neutrino, as the mass is vanishingly
small, the Compton wavelength tends to infinity or turns out to be very
large. On the other hand we encounter the negative energy solutions
within this region. That is we encounter negative energy neutrinoes only.
The equation for a negative energy neutrino is (cf. ref.\cite{r15}).
$$(-p_o)v (p) = + \vec \sigma . \vec p v (p)$$
This is the equation for a left handed neutrino in the physical world of
positive energy solutions.\\
8)There is a close connection between the complex shift of section 7,
equation (\ref{e32}) (and so, ultimately the Kerr-Newman metric), the
hydrodynamical formulation of section 2 and the monopole theory of
section 4.\\
Infact, we could identify $K^\mu$ of section 4 and the momentum vector
$p^\mu$ from section 2 with $\frac{1}{a^\mu}$ of (\ref{e32}). If further $a^\imath$ is taken
to be of the order of the Compton wavelength, $\frac{\hbar}{mc}$ and
similarly $a^o$ to be of the order of $\frac{\hbar}{mc^2}$
we get immideately
$$| \vec p | |\vec a | \sim mc \frac{\hbar}{mc} = \hbar,$$
which can also be obtained from the Heisenberg uncertainity principle.\\
9) It may be remarked that we started in section 3 with purely Quantum
Mechanical postulates and deduced mechanical effects. We came a full circle
in section 8 wherein, from purely classical considerations, we deduced
Quantum Mechanical phenomena.\\
10) The fact that the magnetic field which arises in the monopole
formulation as given by equation (\ref{e22}) and the quantized spin
angular momentum which arises in the hydrodynamical formulation as given
by equation (\ref{e7}) appear to be one and the same is remarkable.
This is caused by the fact that the $\vec K$ and the momentum vector
as given in the two formulations are really one and the same, as pointed
out in section 4. Indeed the Coriolis and other effects
of rotating frames \cite{r39} bear a strong resemblence to the magnetic
effects. As pointed out, electromagnetism and gravitation
can be unified in a general relativistic version of quantum mechanics
as symbolised by the complete description of the electron in terms of the
Kerr-Newman metric. This has been indicated in section 8. Thus in this
picture monopoles disappear. Indeed they have not been found todate and
Dirac himself expressed his conviction that they do not exist
\cite{r40}. (They appear again in the theory of non Abelian
guages.)\\
11) The double valuedness that arises from a nodal singularity on the one
hand and half integral spin on the other finds an immediate echo in the
Kerr-Newman metric. This can be seen as follows.

In natural units the metric is given by (cf. ref.\cite{r24}).
$$ds^2 = - \frac{\Delta}{\rho^2} [dt - a \sin^2 \theta d \phi]^2
+ \frac{\sin^2 \theta}{\rho^2} [(r^2 + a^2) d \phi - a dt]^2
+ \frac{\rho^2}{\Delta} dr^2 + \rho^2 d \theta^2,$$
where, $a$ is the Compton wavelength and,
$$\Delta = r^2 - 2mr + a^2 + m^2 + e^2, \rho^2 \equiv r^2 + a^2 \cos^2 \theta$$
At $r = a$ and $\theta = \pi/2$, $\Delta = 2a^2$ as both $e$ and $m < < a,$
and $\rho^2 = a^2.$\\

If further, we take $a \frac{d\phi}{dt} = \lambda,$ we get,
$$ds^2 = (2\lambda^2 - 1) dt^2 + \frac{1}{2} dr^2$$
The choice $\lambda = \frac{1}{2}$ leads to,
$$ds^2 = - \frac{1}{2} dt^2 + \frac{1}{2} dr^2,$$
which is Minkowski like, except for the scale factor $\frac{1}{\sqrt{2}}.$
In the foregoing model, $a \frac{d\phi}{dt} =$ velocity of light = 1. The
choice $\lambda = \frac{1}{2}$ can be understood as follows: If the
azimuthal angle measured by an observer at rest far away, is $\phi',$ then
we get back the velocity of light at $r = a$ for this observer, if
$\phi' = 2\phi,$ which is precisely spinorial behaviour.

In other words, special relativity for the spin $\frac{1}{2}$ electron can
be seen to emerge from the Kerr-Newman metric.\\
12) More general than the radius of the horizon given in (\ref{e28}) is the
static limit of the Kerr-Newman Black Hole, wherein $'a'$ is replaced by
acos $\theta$, $\theta$ being the usual polar coordinate. However in the
QMBH, as we approach the Compton wavelength $\sim a$, we encounter the
unphysical zitterbewegung region where $\theta$ ceases to have any
physical meaning. In other words, the spin in insensitive to $\theta$
unlike in the classical case. This is ofcourse well known in Quantum
Mechanics.\\
13) As has been pointed out in the introduction, QFT works with point
particles and a space-time continuum in a special relativistic context.
Divergences appear when we go right upto $r = 0$. However, it appears that
for Fermionic Fields atleast, this picture may be valid only for distances
greater than the Compton wavelength: This is in the spirit of Effective
Field Theories \cite{r41,r42}. (In the words of Weinberg, QFT could be a
"low" energy approximation). The above model forbids
such a limiting process for Fermions and sets a cut off. Once we enter the
QMBH region, a very high energy phenomenon, space-time in the conventional
sense becomes unphysical. The appearance of complex coordinates or non-
Hermitian operators is a manifestation of this unphysical feature.\\
14) In the usual formulation of the Hole Theory, the Dirac sea is filled
with negative energy electrons, and by invoking the Pauli exclusion
principle, transitions to negative energy states are forbidden. In the
present formulation, in effect, the Dirac sea of negative energy states is
squeezed into the QMBH and the Quantum Mechanical censor of section 7 forbids
transitions.\\
15) It may be remarked that there have been somewhat similar approaches,
but these do not explain enough or they assume too much, being still
somewhat tentative and preliminary. We discuss some of these very briefly.\\
Barut and Bracken\cite{r43} treat the zitterbewegung effects as a harmonic
oscillator in the Compton wavelength region while spin appears as the
orbital angular momentum associated with the internal system which is
taken to be circulating with velocity $c$ and whose space has a curved
geometry. The rest mass is the internal energy in the rest frame of the
centre of mass of the system. However this model has a number of
shortcomings \cite{r44}.\\
Hestenes \cite{r45} takes a slightly different view treating the
zitterbewegung as arising from self interaction, there being an electro
magnetic wave particle duality, though electron spin is again the orbital
angular momentum with respect to an instantaneous rest system of radius
equalling the Compton wavelength. But a number of assumptions are made
for getting consistency with the Dirac equation.\\
Chacko \cite{r46,r47} following a cue of John A.  Wheeler
models in a somewhat adhoc scheme, elementary particles as superdense
geometrodynamical (that is General Relativisitc) entities confined to
travel with the velocity of light in circular paths, again of radius
equalling the Compton wavelength. Unfortunately properties like spin,
magnetic moment, charge etc. are not incorporated in this scheme.\\
16) A question that arises is, can we arrive at a mass spectrum? A
preliminary indication has been given in
\cite{r48}. The point is that there are a few schemes
which give the mass spectrum of a large number of elementary particles
as composites of pions \cite{r49} or leptons like the electron
positron and neutrino \cite{r50}. Indeed as the discussion following
equation (\ref{e33}) shows, a pion can be considered as an electron
positron composite because its Compton wavelength equals the classical
electron radius which resembles the fact that the pion is a quark anti-
quark composite. These could be incorporated into the above QMBH
considerations to get a mass spectrum\cite{r48}. In particular it is
interesting to note that in the above considerations it is possible to
think of a proton as a composite of two positrons and an electron consistent
with deep inelastic scattering data. Such a
scheme gives a rationale for the matter anti-matter imbalance.\\
17) In recent years the problem of inertial mass has received some
attention, apart from its usual Machian characterisation \cite{r51,r52}.\\
18) Four final comments:\\
In the above considerations regarding the Compton
wavelength we have considered free particles. However as equation
(\ref{e33}) and the subsequent brief discussion indicates the scales could
decrease by a few orders of magnitude in the presence of interaction.\\
For the record it may also be mentioned that the stability and
indivisibility of hydrodynamical vortices had lead to pre-Quantum Mechanical
speculation that atoms may be represented by such vortices\cite{r18}.\\
Effects like the Lamb shift could be explained in the present model by
arguments similar to those in ref.\cite{r44}.\\
The inertial mass of section 3 could also be thought to arise for a free
particle with effective mass $m'$, from the known non-local Schrodinger
equation (cf.ref.\cite{r7}).


\begin{thebibliography}{99}
\bibitem {r1} Vasudevan, R., "Hydrodynamical Formulation of Quantum Mechanics",
in "Perspectives in Theoretical Nuclear Physics", Ed. Srinivasa Rao, K.,
and Satpathy, L., Wiley Eastern, New Delhi, 1994, pp216-225.
\bibitem {r2} Rae, A.I.M., "Quantum Mechanics", IOP Publishing, Bristol,
p222 ff.
\bibitem {r3} Sidharth, B.G., "Quantum Mechanical Black Holes II", Invited
Talk at International Centre of Theoretical Physics, Trieste, Italy. B.M.
Birla Science Centre TR,BSC-CAMCS 96-09-01.
\bibitem {r4} Sidharth, B.G., Nonlinear World (USA), \underline{1}, 1994,
pp403-408.
\bibitem {r5} Sidharth, B.G., "Quantum Mechanical Black Holes", TR BSC-CAMCS-
95-10-01, B.M. Birla Science Centre, Hyderabad.
\bibitem {r6} Feynman, R.P., Leighton, R.B., and Sands, M., "The Feynman
Lectures on Physics", Vol.III, Addison-Wesley Publishing Co., Inc.,
Massachusetts, 1965, Chapter 8 ff.
\bibitem {r7} Mott, N.F., and Massey, H.S.W., "The Theory of Atomic
Collisions", Oxford University Press, London, 1965, p174.
\bibitem {r8} Weinberg, S., "Gravitation and Cosmology", John Wiley \& Sons,
New York, 1972, p.62.
\bibitem {r9} Bjorken, J.D., and Drell, S.D., "Relativistic Quantum Mechanics",
Mc-Graw Hill, New York, 1964, p.39.
\bibitem {r10} Dieks, D., and Nienhius, G., Am.J.Phys., \underline{58} (7),
1990, pp650-655.
\bibitem {r11} Okolowski, J.A., and Slomiana, M., Am.J.Phys., \underline{61}(4),
1993.
\bibitem {r12} Anandan, J., Phys. Rev., \underline{D} 24, 1981, pp.338-346.
\bibitem {r13} Dirac, P.A.M., Proc. Roy. Soc., \underline{A 133}, 1931, pp.60 ff. 1931.
\bibitem {r14} Dirac, P.A.M., "The Principles of Quantum Mechanics",
Clarendon Press, Oxford, 1958, p.263.
\bibitem {r15} Schweber, S.S., "Relativistic Quantum Field Theory", Harper
and Row, New York, 1964, p.47.
\bibitem {r16} Roman, P., "Advanced Quantum Theory", Addison-Wesley,
Reading, Mass, 1965, p.31.
\bibitem {r17} Moller, C., "The Theory of Relativity", Clarendon Press,
Oxford, 1952, pp.170 ff.
\bibitem {r18} Joos, G., "Theoretical Physics", Blackie (London), 1951,
p199ff.
\bibitem {r19} Motz, L., Nuovo Cim., XXXVI (4), 1962.
\bibitem {r20} Motz, L., Nuovo Cim.., 12B (2), 1972.
\bibitem {r21} Markov, M.A., Soviet Phys. JETP, \underline{24}(3),1967, p584.
\bibitem {r22} B.S. De Witt, Phys. Rev, \underline{160}, 1967, p1113.
\bibitem {r23} Fang, L.Z., and Ruffini, R., "Basic Concepts in Relativisitc
Astrophysics", World Scientific, Singapore, 1983, p.123.
\bibitem {r24} Misner, C.W., Thorne, K.S., and Wheeler, J.A., "Gravitation",
W.H. Freeman, San Francisco, 1973, pp 448ff., 543ff.
\bibitem {r25} Newman, E.T., Journal of Math. Phys., (14), 1, 1973, p 102.
\bibitem {r26} Newman, E.T., Enrico Fermi International School of Physics
1975, Proceedings p557.
\bibitem {r27} Lass, H., "Vector and Tensor Analysis", McGraw-Hill Book Co.,
Tokyo, 1950, p295 ff.
\bibitem {r28} Bergmann, P.G., "Introduction to the Theory of Relativity",
Prentice-Hall (New Delhi), 1969, p248ff.
\bibitem {r29} Ferraro, V.C.A., "Electromagnetic Theory", ELBS, London,
1968, p455.
\bibitem {r30} Vonsovsky, S.V., "Magnetism of Elementary Particles", Mir
Publisher, Moscow, 1975, p.28.
\bibitem {r31} Lee, T.D., "Particle Physics and Introduction to Field
Theory", Harwood Academic Publishers, New York, 1981, p.586.
\bibitem {r32} Newton, T.D., and Wigner, E.P., Rev. Mod. Phys., \underline{21}
(3), 1949.
\bibitem {r33} Wignall, J.W.G., Found. Phys. \underline{15}(2), 1985.
\bibitem {r34} Altaisky, M.V. and Sidharth, B.G., "On Quantization of
Self-similar systems - One-particle Excitation States", Int.J. Th.
Phys, \underline{34} (12), 1995. (Also, CAMCS
Tech Report December 1994, B.M. Birla Science Centre).
\bibitem {r35} Rajsekharan., G., "Gauge Theories", in "Gravitation, Quanta
and the Universe", Eds. Prasanna, A.R., Narlikar, J.V., and Vishveswara, C.V.,
Wiley Eastern, New Delhi, 1980, pp.208 ff.
\bibitem {r36} Sidharth, B.G., and Popova, A.D., Non Linear World (USA),
\underline{3}, 1996.
\bibitem {r37} Sidharth, B.G., and Popova, A.D., Differential Equations and
Dynamical Systems (DEDS), \underline{4}(3/4),1996, pp431-440.
\bibitem {r38} Ahranov, Y., and Bohm, D., Phys. Rev., 1959, \underline{115},
p.485. (Current Science, \underline{66} (10), 25 May 1994 carries a review
article).
\bibitem {r39} Vladimirov, Yu., Mitskievich, N., and Horsky, J., "Space,
Time,Gravitation", Mir Publishers, Moscow, 1987, p.87ff.
\bibitem {r40} Craigie, N.S., Goddard, P., and Nahm, W., Eds., "Monopoles in
Quantum Field Theory", World Scientific, Singapore, 1982.
\bibitem {r41} Sidharth, B.G., "Quantum Mechanical Non Linearities in the
context of Black Holes", Nonlinear World (USA), \underline{4}, 1997.
\bibitem {r42} Davies, P., (Ed.) "The New Physics", Cambridge University
Press (Cambridge), 1989, p 446ff gives a semi technical discussion of
Effective Field Theories.
\bibitem {r43} Barut, A.O., and Bracken, A.J., Phys. Rev. D 23, (10), 1981.
\bibitem {r44} Hestenes, D., Found of Phys, 15(1), 1985.
\bibitem {r45} Hestenes, D., Found of Phys, 20(10), 1990.
\bibitem {r46} Chacko, T., Int. J. Th. Phys 11(1), 1974.
\bibitem {r47} Chacko, T., Int. J. Th. Phys 12(2), 1975.
\bibitem {r48} Sidharth, B.G., "Quarks, Leptons and a Mass Spectrum", TR,
BSC-CAMCS 96-10-02b (Also paper in DAE (Govt. of India) Symposium on
Nuclear Physics, 1996.)
\bibitem {r49} Kenny, J., IJTP, \underline{13} (5), 1975.
\bibitem {r50} Barut, A.O., Surv. High Energy Phys. 1(2), 113, 1980.
\bibitem {r51} Assis, A.K.T., Found. Phys. Lett., 2, 1989, pp 301-318.
\bibitem {r52} Haisch, B., Rueda, A., and Puthoff, H.E., Phys. Rev.
\underline{A49}(2), 1994, pp 678-694.
\end{thebibliography}
\end{document}